\documentclass[unsortedadress,preprint
 ,secnumarabic%
,amssymb, amsmath,nobibnotes,prd]{revtex4}
\begin{document}
\title{Berezin Quantization of Gauged WZW and Coset Models\footnote{This
work is dedicated to
Professor Michel Ryan on the occasion of his 60th birthday.}}

\author{I. Carrillo-Ibarra}
\affiliation{Departamento de Matem\'aticas\\
Centro de Investigaci\'on y de Estudios Avanzados del IPN\\
P.O. Box 14-740, 07000 M\'exico D.F., M\'exico}
\author{H. Garc\'{\i}a-Compe\'an\email{\tt
compean@fis.cinvestav.mx} and 
W. Herrera-Su\'arez}
\affiliation{Departamento de F\'{\i}sica\\
Centro de Investigaci\'on y de Estudios Avanzados del IPN\\
P.O. Box 14-740, 07000 M\'exico D.F., M\'exico}

\vskip 1truecm
\date{\today}

\begin{abstract} 
Gauged WZW and coset models are known to be useful to prove holomorphic
factorization of the partition function of WZW and coset models. In this note we show that 
these gauged models can be also important to quantize the theory
in the context of the Berezin formalism. For gauged coset models Berezin quantization procedure also 
admits a further holomorphic factorization in the complex structure of the moduli space. 
\end{abstract}

\vskip -1truecm
\maketitle

\vskip -1.3truecm
\newpage

\setcounter{equation}{0}

\section{Introduction}

The application of diverse quantization methods to physical systems gives, in many cases,
complementary information about these systems. Wess-Zumino-Witten (WZW) models are very
interesting kind of two dimensional conformal field theories (CFT) representing exactly solvable models which have been studied
intensively some years ago. These models have been studied in the context of different
quantization procedures as canonical quantization \cite{cq} and Feynman path integral (for
a review see, for instance, \cite{mblau,hori} and references therein). 

The coupling of WZW and coset models to gauge fields
constitutes the gauged
WZW and coset models \cite{kupianen}. In the context of nonsupersymmetric theories these models have been
discussed in various global contexts in \cite{mblau,hori,wzw,grassman,stanciu}. In
particular, in
Ref.
\cite{wzw}, Witten have used them to give an alternative proof of the holomorphic factorization
of the
partition function of WZW and coset models. In this proof the methods of geometric
quantization and its relation to the canonical quantization of Chern-Simons theory \cite{cs}
have been very useful. In the present note we use some of Witten results of Ref. \cite{wzw}, to quantize gauged WZW
and coset models through the Berezin formalism \cite{berezin} (for recent developments, see
for instance, 
\cite{rcgone,cahen,kara,martinone,martinthree,takhtajan,maciej,isidro,ilis} to carry over this procedure. In particular, 
we will use some of the results obtained in \cite{ilis}. Through out this
paper we follow the notation and
conventions given by Witten in Ref. \cite{wzw}.

To begin with we first recall the structure of the WZW model described by the Lagrangian
\begin{equation}
L(g) = -{k \over 8 \pi} \int_{\Sigma} d^2 \sigma \sqrt{\rho} \rho^{ij} {\rm Tr}
\big( g^{-1} \partial_i g \cdot g^{-1} \partial_j g \big) - {ik\over 12 \pi} \int_M d^3
\sigma \varepsilon^{ijk} {\rm Tr} \big( g^{-1} \partial_i
g \cdot g^{-1} \partial_j g \cdot g^{-1} \partial_k g \big), 
\end{equation}
where $k \in \mathbb{Z}$ is the level, $\rho_{ij}$ is the worldsheet metric, $g: \Sigma \to G$ 
is a map of a
compact and orientable Riemann surface $\Sigma$ (without boundary) into a simple, compact, 
connected and simply connected Lie group $G$ and $M$ is a three-dimensional manifold whose boundary $\partial M$ is $\Sigma$. The partition function of the WZW is defined as $Z_{WZW}(\Sigma) = \int {\cal D}g e^{-L(g)}.$

The WZW Lagrangian $L(g)$ is invariant under the {\it global} action of $G_L \times G_R$ on
$G$ given by $g \to agb^{-1}$ with $a\in G_L$ and $b \in G_R$. Here $G_L$ and $G_R$ are copies of $G$. However if
one gauge out
a subgroup $F$ of $G_L \times G_R$, for instance $F= G_R$. This WZW action gives rise to the coupling between the fields $g$ and $F$ gauge fields on $\Sigma$ through the following action 
\begin{equation}
L(g,A) = L(g) + {k\over 2 \pi} \int_{\Sigma} d^2 z  {\rm Tr} A_{\bar{z}}
g^{-1}
\partial_z g - {k \over 4 \pi} \int_{\Sigma}  d^2 z  {\rm Tr} A_{z} A_{\bar{z}},
\end{equation}
which is not, in general, a gauge invariant extension. The map $g$ is now generalized to a section $g \in \Gamma(\Sigma,X)$ of the $F$-bundle over $\Sigma$: $F \to X \buildrel{\pi}\over{\to} \Sigma$. The introduced gauge connection is a
${\bf f}$-valued connection one-form on $X$, transforming in the {\it adjoint} representation of the gauge group
$F$, where ${\bf f}$ is the Lie algebra of $F$. Only for special ``anomaly free'' gauge
groups $F$ there exists such a gauge invariant extension. But we will consider, in the present paper, `anomalous' gauge groups $F$'s such that under the infinitesimal gauge transformation
\begin{equation}
\delta g = - g u, \  \  \ \ \ \ \  \delta A_i = - D_i u = - \partial_i u - ´[A_i,u],
\end{equation} 
$L(g,A)$ differs from zero in the form
\begin{equation}
\delta L(g,A) = {ik \over 4
\pi}\int_{\Sigma} d^2 z  {\rm Tr} u dA.
\end{equation}

Following Witten \cite{wzw}, one can define the functional
$$
\psi(A) = \int {\cal D} g e^{-L(g,A)} 
$$
\begin{equation}
= \int {\cal D} g \exp\bigg( -L(g) - {k \over 2 \pi}\int_{\Sigma} d^2 z {\rm Tr}
A_{\bar{z}} g^{-1}
\partial_z g + {k \over 4 \pi} \int_{\Sigma}  d^2 z  {\rm Tr} A_{\bar{z}} A_{z} \bigg).
\end{equation}
This functional obeys the following equations
\begin{equation}
\bigg({\delta \over \delta A_z} - {k \over 4 \pi}A_{\bar{z}}\bigg) \psi
(A) =0,
\label{holoone}
\end{equation}
and
\begin{equation}
\bigg( D_{\bar{z}} {\delta \over \delta A_{\bar{z}}} + {k \over 4 \pi} D_{\bar{z}} A_z
-{k \over 2 \pi} F_{{\bar{z}}z} \bigg) \psi(A) =0,
\label{gaugeinv}
\end{equation}
where $F_{\bar{z} z} = \partial_{\bar{z}} A_z - \partial_z A_{\bar{z}}.$
If one define the operators: ${D \over DA_z} = {\delta \over \delta A_z} - {k \over 4 \pi}
A_{\bar{z}}$ and ${D \over DA_{\bar{z}}} =
{\delta \over \delta A_{\bar{z}}} + {k \over 4 \pi} A_{z},$
we can rewrite Eqs. (\ref{holoone}) and (\ref{gaugeinv}) as
\begin{equation}
{D \over DA_z} \psi(A)=0,
\label{holo}
\end{equation}
and
\begin{equation}
\bigg(D_{\bar{z}} {D \over DA_{\bar{z}}} -  {k \over 2 \pi}  F_{{\bar{z}}z}\bigg) \psi(A) =0.
\label{gauge}
\end{equation}

${D \over DA_z}$ and ${D \over DA_{\bar{z}}}$ can be regarded as a gauge connection on
a unitary line bundle ${\cal L}^{\otimes k}$ over the  space of all connections ${\cal
A}$ over $\Sigma$. The curvature of this connection can be computed by using the quantization condition:
$\big[ {D \over DA_z(z)}, {D \over DA_{\bar{w}}(w)}\big] = {k \over 2 \pi} \delta(z,w)$
and it yields $-i \omega$, where  
$\omega = k \omega_0$ is the symplectic form on ${\cal A}$ with 
$\omega_0 = {1 \over 2 \pi} \int_{\Sigma} {\rm Tr} \delta A \wedge
\delta A$. Then $\omega$ gives to ${\cal A}$ the structure of a symplectic manifold $({\cal A}, \omega)$ and this suggest the geometric prequantization of
${\cal A}$. As ${\cal A}$ is topologically trivial, the prequantum line bundle ${\cal L}^{\otimes k}$ can
be identified with the {\it trivial} holomorphic line bundle: $ {\cal P}= {\cal A} \times \mathbb{C}$, 
whose $L^2$-completion of holomorphic sections constitutes the Hilbert space represented by $H^0_{L^2}({\cal A}, {\cal L}^{\otimes k})$ with Hermitian inner product 
\begin{equation}
\langle \chi|\psi \rangle_{\cal L} = {1 \over {\rm vol}({F})}\int_{{\cal A}} {\cal
D}A \overline{\chi}(A) {\psi}(A). 
\label{innerno}
\end{equation}
The measure ${\cal D}A$ on ${\cal A}$ can be determined by the symplectic structure $\omega$ and it can be written
as
\begin{equation}
\langle \chi|\psi \rangle_{\cal L} = {1 \over {\rm vol}({F})}\int_{{\cal A}} \overline{\chi}(A) {\psi}(A)
\exp\big(-\Phi\big) {\omega^n \over n!}, 
\label{inner}
\end{equation}
where we have divided by the volume of the gauge group $F=G$ and where $\Phi$ is the K\"ahler
potential of the metric on ${\cal A}$. Many of these results about the differential geometry of
this moduli space were firstly described in Ref. \cite{abott}.

The curvature of the connection compatible with the Hermitian structure is
given by $\bar{\partial} \partial (-\Phi) = -i \omega$. Of course the existence of a
prequantization bundle implies that $\big[{\omega \over 2 \pi}\big] \in H^2({\cal A}_J,
\mathbb{Z})$. If one picks a complex structure on $\Sigma$ it induces on ${\cal A}$ a {\it fixed} complex structure $J$ giving rise to a complex K\"ahler manifold ${\cal A}_J$. Complex structure $J$ also induces a K\"ahler polarization on ${\cal L}^{\otimes k}$ which completes the {\it geometric quantization} of the K\"ahler manifold ${\cal A}_J$. 

The prequantum line bundle can be pushed-down as follows. The symplectic action of the gauge group 
$\widehat{F}_{\mathbb{C}}$ defined by: $\widehat{F}_{\mathbb{C}}:= \{ f: \Sigma \to {F}_{\mathbb{C}} \}$ (here 
${F}_{\mathbb{C}}$ is the complexification of the gauge group $F$) on ${\cal A}_J$ can be lifted in such a way that it preserves
the connection and the Hermitian inner product (\ref{innerno}) or (\ref{inner}). One may define the pushdown
line bundle $\widetilde{\cal L}^{\otimes k}$ by stating that its sections 
$H^0_{L^2}({\cal M}_J, \widetilde{\cal L}^{\otimes k})= H^0_{L^2}(F^{-1}_J(0), {\cal L}^{\otimes k})^{\widehat{F}_{\mathbb{C}}}$, constitutes a $\widehat{F}_{\mathbb{C}}$-invariant subspace of the space of sections $\Gamma({\cal A}_J,{\cal L})$. Here ${\cal M}_J=F^{-1}_J(0)//\widehat{F}_{\mathbb{C}}$ is the Marsden-Weinstein quotient with 
$F^{-1}_J(0) \subset {\cal A}_J$ is an $\widehat{F}_{\mathbb{C}}$-invariant subspace, where $F_J:{\cal A}_J \to {\bf f}^{*}_{\mathbb{C}}$ is the moment map, with ${\bf f}^{*}_{\mathbb{C}}$ being the dual of the Lie algebra ${\bf f}_{\mathbb{C}}$ of $\widehat{F}_{\mathbb{C}}$. Equations (\ref{holo}) and (\ref{gauge}) implies that the wave function $\psi(A)$ is a
holomorphic and gauge invariant section and therefore it belongs to $H^0_{L^2}({\cal M}_J, \widetilde{\cal L}^{\otimes k}).$ The connection also can be pushed-down and it satisfies $\widetilde{\nabla}_{\widetilde{V}}
\psi = {\nabla}_{{V}} \psi .$ Meanwhile the curvature of the connection
$\widetilde{\nabla}$ is $-i \widetilde{\omega}$. Thus the pushed-down prequantization is given by
$(\widetilde{\cal L}, \widetilde{\nabla}, \langle \cdot | \cdot
\rangle_{\widetilde{\cal L}})$, where $\langle \cdot | \cdot \rangle_{\widetilde{\cal
L}}$ is the $\widehat{F}_{\mathbb{C}}$-invariant inner product $\langle \cdot | \cdot \rangle^{\widehat{F}_{\mathbb{C}}}_{\cal L}$. 

If one {\it varies} the complex structure over the space ${\cal Z}$ of all conformal classes of worldsheet metrics ${\rho}$. Then one can construct the following vector bundle:
$H^0_{L^2}({\cal A}_J, {\cal L}^{\otimes k}) - {\cal V} \buildrel{\pi}\over {\to} {\cal Z}.$
This bundle can be also pushed-down to the K\"ahler quotient ${\cal M}_J$ such that one get:
\begin{equation}
H^0_{L^2}({\cal M}_J, \widetilde{\cal L}^{\otimes k}) - \widetilde{\cal V} \buildrel{\widetilde{\pi}}\over {\to} \widetilde{\cal Z}.
\label{bundle}
\end{equation}
This bundle admits a projectively flat connection which helps to show that the geometric quantization procedure is independent on the complex structure \cite{cs}. A covariantly constant section of the bundle (\ref{bundle}) in an arbitrary basis $\{ {\bf e}_{\ell}\}$ where $\ell=1, \dots , {\rm dim} (\widetilde{\cal V}')$ (with 
$\widetilde{\cal V}' = \widetilde{\cal V}'_{\rho} = H^0_{L^2}({\cal M}_J, \widetilde{\cal L}^{\otimes k})$) can be written as:
\begin{equation}
\psi(A;\rho) = \sum_{\ell=1}^{ {\rm dim} \widetilde{\cal V}'} {\bf e}_{\ell}(A;\rho) \cdot \overline{f}_{\ell}(\rho),
\end{equation}
for some expansion coefficients $\overline{f}_{\ell}(\rho)$ which are standard anti-holomorphic functions on $\widetilde{\cal Z}$. The partition function of WZW models $Z_{WZW}(\Sigma;\rho)$ can be expressed in terms of these coefficients as follows \cite{wzw}
\begin{equation}
Z_{WZW}(\Sigma;\rho) = \sum_{\ell=1}^{ {\rm dim} \widetilde{\cal V}'} |{f}_{\ell}(\rho)|^2.
\end{equation}

\section{Berezin Quantization of Gauged WZW Models}

In this section we describe the Berezin's quantization of the K\"ahler quotient $({\cal M}_J,
\widetilde{\omega})$ where ${\cal M}_J={F}^{-1}_J(0)/\widehat{F}_{\mathbb{C}}$.  That means we
will find an associative and noncommutative family of algebras $(\widetilde{{\cal S}_B},
\widetilde{*_B})$ with $\widetilde{{\cal S}_B} \subset C^{\infty}({\cal M}_J)$ being the space
of {\it covariant symbols} (which are indexed with a real and positive parameter $\hbar$ in
order to recover the classical limit $\hbar \to 0$) and $\widetilde{*_B}$ is the Berezin star product. In 
order to do that we first {\it Berezin quantize} $({\cal A}_J, \omega)$
and then project out all relevant quantities to be $\widehat{F}_{\mathbb{C}}$-invariant one
finally get $(\widetilde{\cal S}_B, \widetilde{*_B})$ \cite{ilis,memoirs}, where $\widetilde{{\cal S}_B} \subset C^{\infty}({\cal A}_J)^{\widehat{F}_{\mathbb{C}}} \equiv C^{\infty}(F^{-1}_J(0)/{\widehat{F}_{\mathbb{C}}}).$
 
Consider a given prequantization $(\widetilde{\cal L}, \widetilde{\nabla}, \langle \cdot |
\cdot\rangle_{\widetilde{\cal L}})$ of the K\"ahler manifold ${\cal M}_J$, which can be regarded as the  pushed-down prequantization with $\widetilde{\cal L} = {\cal L}^{\widehat{F}_{\mathbb{C}}}$ being the
$\widehat{F}_{\mathbb{C}}$-invariant unitary line bundle over ${\cal A}_J.$ The inner product $\langle \cdot |
\cdot\rangle_{\widetilde{\cal L}}$ is the $\widehat{F}_{\mathbb{C}}$-invariant inner product compatible with the
connection $\widetilde{\nabla}$ constructed from $\langle \cdot | \cdot\rangle_{{\cal L}}$ and it
is given by 

\begin{equation}
\langle \widetilde{\chi} | \widetilde{\psi} \rangle_{\widetilde{{\cal L}}} = 
 \int_{{\cal M}_J} \langle \widetilde{\chi} | \widetilde{\psi} \rangle
{\widetilde{\omega} \over n!} =
\langle \chi | \psi
\rangle^{\widehat{F}_{\mathbb{C}}}_{\cal L} = \langle {\chi} | {\psi} \rangle_{{{\cal
L}}}, 
\end {equation}
where $\langle \chi|\psi \rangle_{\cal L}$ given by Eqs. (\ref{innerno}) or (\ref{inner}). This inner product is defined for all $\widetilde{\chi},\widetilde{\psi} \in H^0_{L^2}({\cal M}_J,\widetilde{\cal L}^{\otimes 
k}) =
H^0_{L^2}(F^{-1}_J(0),{\cal L}^{\otimes k})^{\widehat{F}_{\mathbb{C}}}$
where $\langle \widetilde{\chi} | \widetilde{\psi} \rangle = \langle \chi | \psi \rangle$
with $\langle \chi |
\psi \rangle = \exp \big( -\Phi \big) \overline{\chi} \psi$ and $\Phi$ is the K\"ahler
potential. Also
$\widetilde{\omega}$ is preserved by the action of $\widehat{F}_{\mathbb{C}}$, {\it i.e.} $\omega$ is $\widehat{F}_{\mathbb{C}}$-invariant. The norm of an element
$\widetilde{\psi}$ of $H^0_{L^2}(F^{-1}_J(0),{\cal L}^{\otimes k})^{\widehat{F}_{\mathbb{C}}}$ is defined by 
$ [||\widetilde{\psi}
||^2]_{\widetilde{\cal L}} \equiv 
\langle \widetilde{\psi} |
\widetilde{\psi}
\rangle_{\widetilde{\cal L}} =||\psi ||^2_{\cal L}.$

In Ref. \cite{wzw} it was shown that one can identify the complex conjugate $\overline{\psi}(A)$ of ${\psi}(A)$ 
with $\chi(A)$, if $\chi(A)$ is defined by $\chi(A)= \int {\cal D}h e^{-L'(h,A)}$, where $L'(h,A)$ is given by
\begin{equation}
L'(h,A) = L(h) - {k\over 2 \pi} \int_{\Sigma} d^2 z  {\rm Tr} A_{z}
\partial_{\bar{z}} h h^{-1}   - {k \over 4 \pi} \int_{\Sigma}  d^2 z  {\rm Tr}   A_{\bar{z}}
A_{z}.
\end{equation}
The computation of the norm $[||\widetilde{\psi}
||^2]_{\widetilde{\cal L}}=||\psi ||^2_{\cal L} = {1 \over {\rm vol}(\widehat{F}_{\mathbb{C}})}\int_{{\cal A}_J} {\cal
D}A \overline{\psi}(A) {\psi}(A)$ by integrating out with respect to $A$ (with an appropriate regularization procedure) and the uses of the formula of Polyakov and Wiegman ensures the holomorphic factorization of the partition function of the WZW model, {\it i.e.} $Z_{WZW}(\Sigma) = ||\psi(A) ||^2_{\cal L}.$

For future reference, we proceed to give a global set up for the Berezin quantization \cite{rcgone,cahen,kara}. We take $\widetilde{\cal Q} \in
\widetilde{\cal L}_0^{\otimes k},$ $\pi[\widetilde{\cal Q}] = \widetilde{A} \in {\cal M}_J$, where  $\widetilde{\cal L}_0^{\otimes k}$
is a complex unitary line bundle
$\widetilde{\cal L}^{\otimes k}$ without the zero section. Now
consider $\widetilde{\psi}(\widetilde{A}) = \widetilde{\psi}[\pi(\widetilde{\cal Q})] =
\widetilde{L_{\cal Q}} [\widetilde{\psi}]
\widetilde{\cal Q}$ with  $\widetilde{L_{\cal Q}} [\widetilde{\psi}]$ being a linear functional of
$\widetilde{\psi}$. The group $\widehat{F}_{\mathbb{C}}$ acts on the space of sections $H^0_{L^2}(F^{-1}_J(0),{\cal L}_0^{\otimes k})^{\widehat{F}_{\mathbb{C}}}$
in the
equivariant form $(\widetilde{\Gamma} \widetilde{\psi})(\widetilde{A}) \equiv \widetilde{\Gamma} \widetilde{\psi}
(\widetilde{\Gamma}^{-1} \widetilde{A}),$ 
where $\widetilde{\Gamma} \in \widehat{F}_{\mathbb{C}}$, $\widetilde{A} \in {\cal M}_J$ and $\widetilde{\psi}
\in
H^0_{L^2}(F^{-1}_J(0),{\cal L}_0^{\otimes k})^{\widehat{F}_{\mathbb{C}}}.$

On the other hand Riesz theorem implies the existence of a section $\widetilde{e_{\cal Q}} \in
H^0_{L^2}(F^{-1}_J(0),{\cal L}^{\otimes k}_0)^{\widehat{F}_{\mathbb{C}}}$ such
that $\widetilde{L_{\cal Q}} [\widetilde{\psi}] = \langle \widetilde{e_{\cal Q}}| \widetilde{\psi}
\rangle_{\widetilde{\cal L}_0}$. $\widetilde{e_{\cal Q}}$ is a $\widehat{F}_{\mathbb{C}}$-invariant 
section called the equivariant generalized coherent state. 

Now let $\widehat{\cal O}^{\widehat{F}_{\mathbb{C}}} : H^0_{L^2}(F^{-1}_J(0), {\cal L}^{\otimes k}_0)^{\widehat{F}_{\mathbb{C}}} \to
H^0_{L^2}(F^{-1}_J(0), {\cal L}^{\otimes k}_0)^{\widehat{F}_{\mathbb{C}}}$ be a {\it bounded} operator. The {\it
covariant symbol} of
this operator is defined as
\begin{equation}
{\cal O}_B^{\widehat{F}_{\mathbb{C}}}(\widetilde{A}) = { \langle \widetilde{e_{\cal Q}}|
\widehat{\cal O}^{\widehat{F}_{\mathbb{C}}}| \widetilde{e_{\cal Q}} \rangle_{\widetilde{\cal L}_0} \over [||
\widetilde{e_{\cal Q}} ||^2]_{\widetilde{\cal L}_0}} \equiv { \langle e_{\cal Q}| \widehat{\cal O}|
e_{\cal Q} \rangle_{{\cal L}_0}^{\widehat{F}_{\mathbb{C}}} 
\over [|| e_{\cal Q} ||^2]_{{\cal L}_0}^{\widehat{F}_{\mathbb{C}}}},
\end{equation}
where $|| \widetilde{e}_{\cal Q}||^2_{\widetilde{\cal L}_0}$ is given by
\begin{equation}
[|| \widetilde{e}_{\cal Q} ||^2]_{\widetilde{\cal L}} = [||{e}_{\cal Q} ||^2]_{\cal L}^{\widehat{F}_{\mathbb{C}}} =
 {1 \over {\rm vol}(\widehat{F}_{\mathbb{C}})}\int_{{\cal M}_J} {\cal
D}\widetilde{A} \overline{\widetilde{e_{\cal Q}}}(\widetilde{A}) {\widetilde{e_{\cal Q}}}(\widetilde{A}),
\end{equation}
where $e_{\cal Q} \in H^0_{L^2}(F^{-1}_J(0), {\cal L}^{\otimes k}_0)$ and $\widehat{\cal O}: H^0_{L^2}(F^{-1}_J(0), {\cal L}^{\otimes k}_0) \to H^0_{L^2}(F^{-1}_J(0), {\cal L}^{\otimes k}_0)$ is a bounded operator.

The space of covariant symbols $\widetilde{{\cal S}_B}={\cal S}_B^{\widehat{F}_{\mathbb{C}}}$ is defined as the pushing-down of ${\cal
S}_B$. Each covariant symbol can be analytically continued to the open dense subset of
${\cal M}_J \times {\cal M}_J$ in such a way
$\langle \widetilde{e_{{\cal Q}}}| \widetilde{e_{{\cal Q}'}}\rangle_{\widetilde{\cal L}} \not=
0$ with $\pi(\widetilde{\cal Q}) = \widetilde{A}$ (with
local complex coordinates $\{\widetilde{A}_z,\widetilde{A}_{\bar{z}}\}$) and  $\pi(\widetilde{{\cal Q}'})
= \widetilde{A'}$ (with local complex coordinates $\{\widetilde{A}'_w,\widetilde{A}'_{\bar{w}}\}$), which is holomorphic in the first entry and anti-holomorphic in the second one. This analytic
continuation is reflected in the covariant symbol in the form
\begin{equation}
{\cal O}_B^{\widehat{F}_{\mathbb{C}}}(\widehat{A}_{\overline{z}},\widehat{A}'_{{w}}) = {\langle e_{\cal Q}| \widehat{\cal O}| e_{{\cal
Q}'}\rangle_{{\cal L}_0}^{\widehat{F}_{\mathbb{C}}} \over
\langle e_{\cal Q} | e_{{\cal Q}'}\rangle_{{\cal L}_0}^{\widehat{F}_{\mathbb{C}}}}.
\end{equation}

The operator $\widehat{\cal O}^{\widehat{F}_{\mathbb{C}}}$ can be obtained from its symbol in the form
\begin{equation}
\widehat{\cal O}^{\widehat{F}_{\mathbb{C}}} \widetilde{\psi} (\widetilde{A}_{\overline{z}}) = \langle e_{{\cal Q}}| \widehat{\cal O}| \psi 
\rangle_{{\cal L}_0}^{\widehat{F}_{\mathbb{C}}} {\cal Q}.
\label{aplica}
\end{equation}
The consideration of the completeness condition $ {\bf 1} = \int_{{\cal A}_J} | e_{\cal Q}
\rangle \langle e_{\cal Q}| \exp \big(-\Phi(A_z,A_{\bar{z}}) \big) {\omega^n \over
n!}(A_z,A_{\bar{z}})$ in the computation of $\langle e_{{\cal Q}}| \widehat{\cal O}| \psi 
\rangle_{{\cal L}_0}  {\cal Q}$ yields
\begin{equation}
\widehat{\cal O} \psi (A_{\overline{z}}) = \int_{{\cal A}_J} {\cal D} {A}'
{\cal O}_B(A_{\overline{z}},A'_{{w}}){\cal B}_{\cal Q}(A_{\overline{z}},A'_{{w}})
\psi
(A'_w) \ {\cal Q},
\end{equation}
or in terms of the symplectic structure we have
\begin{equation}
\widehat{\cal O} \psi (A_{\overline{z}}) = \int_{{\cal A}_J} {\cal O}_B(A_{\overline{z}},A'_{{w}}){\cal B}_{\cal Q}(A_{\overline{z}},A'_{{w}})
\psi
(A'_w) \exp \bigg( - \Phi(A'_{\overline{w}},A'_{{w}}) \bigg)
{\omega^n \over n!}   
(A'_{\overline{w}},A'_{{w}}) {\cal Q},
\end{equation}
where $\psi(A'_w) = \langle e_{{\cal Q}'} | \psi \rangle_{{\cal L}_0}$ and ${\cal B}_{\cal Q}(A_{\overline{z}},A'_{{w}}) \equiv 
\langle e_{{\cal Q}} | e_{{\cal Q}'}\rangle_{{\cal L}_0}.$ ${\cal B}_{\cal
Q}(A_{\overline{z}},A'_{{w}})$ is the  {\it generalized Bergman kernel}. Finally, taking the 
$\widehat{F}_{\mathbb{C}}$-invariant part of the above expression we get Eq. (\ref{aplica}).

Similar considerations apply to other formulas. But an essential difference with respect to the
quantization of $({\cal A}_J,\omega)$ is that, in the present case, the K\"ahler quotient is
topologically nontrivial and therefore the line bundle $\widetilde{\cal L}^{\otimes k}$ is also {\it
nontrivial}. It is only locally trivial {\it i.e.} $\widetilde{\cal L}^{\otimes k}_{(j)} = {\cal W}^{(j)}
\times \mathbb{C}$ for each dense open subset ${\cal W}^{(j)}_J \subset {\cal M}_J$ with $j =1,2,
\dots, N$. Analog global formulas found on ${\cal L}$, can be applied only on each local
trivialization of $\widetilde{\cal L}^{\otimes k}$. Of course, transition functions on ${\cal W}^{(i)}_J
\cap {\cal W}^{(j)}_J$ with $i \not= j$ are very important and {\it sections} and other relevant
quantities like the {\it Bergman kernel}, {\it K\"ahler potential}, {\it covariant symbols},
etc., transform nicely under the change of the open set. Thus in a particular
trivialization $\widetilde{\cal L}^{\otimes k}_{(j)}$, the function ${\cal
O}^{(j)}_{B(0)}(\widetilde{A}_{\overline{z}},\widetilde{A}'_{{w}}) \in C^{\infty}({\cal W}^{(j)}_J)$ is called the {\it covariant
symbol} of the operator $\widehat{O}^{(j)}_0$ acting on $H^0_{L^2}({\cal W}^{(j)}_J, \widetilde{\cal L}^{\otimes k}_{(j)}).$ Now if ${\cal O}^{(j)}_{B(0)}(\widetilde{A}_{\overline{z}},\widetilde{A}'_{{w}})$
and ${\cal O}'^{(j)}_{B(0)}(\widetilde{A}'_{\overline{w}},\widetilde{A}'_z)$ are two covariant symbols associated to $\widehat{\cal
O}^{(j)}_0$ and $\widehat{\cal O}'^{(j)}_0$, respectively, then the covariant symbol of
$\widehat{{\cal O}}^{(j)}_0\widehat{\cal O}'^{(j)}_0$ is given by the {\it Berezin-Wick star
product} ${\cal O}^{(j)}_{B(0)} {*_B} {\cal O}'^{(j)}_{B(0)}$
$$
({\cal O}^{(j)}_{B(0)} \widetilde{*_B} {\cal O}'^{(j)}_{B(0)})(\widetilde{A}_{\overline{z}},\widetilde{A}_{{z}})
$$
$$
=\int_{{\cal W}^{(j)}_J} {\cal O}^{(j)}_{B(0)}(\widetilde{A}_{\overline{z}},\widetilde{A}'_{{w}}) {\cal
O}'^{(j)}_{B(0)}(\widetilde{A}'_{\overline{w}},\widetilde{A}_{{z}})
{{\cal
B}^{(j)}(\widetilde{A}_{\overline{z}},\widetilde{A}'_{{w}}){\cal B}^{(j)}(\widetilde{A}'_{\overline{w}},\widetilde{A}_{{z}}) \over {\cal
B}^{(j)}(\widetilde{A}_{\overline{z}},\widetilde{A}_{{z}})}
\exp \bigg\{- \Phi^{(j)}(\widetilde{A}'_{\overline{w}},\widetilde{A}'_{{w}}) \bigg\} {\widetilde{\omega} \over
n!}(\widetilde{A}'_{\overline{w}},\widetilde{A}'_{{w}})
$$
\begin{equation}
= \int_{{\cal W}^{(j)}_J} {\cal O}^{(j)}_{B(0)}(\widetilde{A}_{\overline{z}},\widetilde{A}'_{{w}}) {\cal
O}'^{(j)}_{B(0)}(\widetilde{A}'_{\overline{w}},\widetilde{A}_{{z}})
\exp \bigg\{
{\cal K}^{(j)}(\widetilde{A}_{\overline{z}},\widetilde{A}_{{z}}|\widetilde{A}'_{\overline{w}},\widetilde{A}'_{{w}}) \bigg \} {\widetilde{\omega} \over n!}
(\widetilde{A}'_{\overline{w}},\widetilde{A}'_{{w}}),
\end{equation}
where ${\cal K}^{(j)}(\widetilde{A}_{\overline{z}},\widetilde{A}_{{z}}|\widetilde{A}'_{\overline{w}},\widetilde{A}'_{{w}}):=
\Phi^{(j)}(\widetilde{A}_{\overline{z}},\widetilde{A}'_{{w}})+
\Phi^{(j)}(\widetilde{A}'_{\overline{w}},\widetilde{A}_{{z}})-
\Phi^{(j)}(\widetilde{A}_{\overline{z}},\widetilde{A}_{{z}})- \Phi^{(j)}(\widetilde{A}'_{\overline{w}},\widetilde{A}'_{{w}})$ is called the {\it Calabi
diastatic
function} on ${\cal W}^{(j)}_J$. This construction is valid for all local prequantizations:
$(\widetilde{\cal L}_{(j)}^{\otimes k}, \widetilde{\nabla}^{(j)}, \langle \cdot |
\cdot\rangle_{\widetilde{\cal
L}_{(j)}})$ with $j=1, \dots , N$. Finally, this structure given by the pair $(\widetilde{{\cal S}_B}, \widetilde{*_B})$
constitutes the Berezin
quantization of $({\cal M}_J,\widetilde{\omega})$ which is determined by the $\widehat{F}_{\mathbb{C}}$-gauged WZW model. 

\section{Berezin Quantization of Coset Models}

\subsection{The $G/H$ Model}

Coset models $G/H$ are CFT's which are equivalent to the gauged WZW models by gauging an anomaly-free subgroup 
$F$ of $G_L \times G_R$ \cite{kupianen}. For instance consider any subgroup $H \subset G_{adj}$ with $G_{adj}$
being the diagonal subgroup of $G_L \times G_R$. In this case $H$ is always anomaly-free. 
 
We now consider the case of a subgroup $F\subset G_L \times G_R,$ which is not anomaly-free. In addition we take $H
\subset G_L$, such that we have $F=G_R \times H_L$. Then an $F$ connection consist of a pair $(A,B)$ of two connections: one $H$-valued connection $B$ and a $G$-valued connection $A$. If $G$ has a non-trivial center $Z(G)$ (diagonally embedded in $G_L \times G_R$), then the symmetry group is  $G_L \times G_R/Z(G)$. Therefore the subgroup that acts faithfully is not $F=G_R \times H_L$ but $F'= G_R \times H_L/Z$, where $Z=H \cap Z(G)$.

Similarly to the case of gauged WZW model we can consider the case of an `anomalous' extension of the $G/H$ model called gauged coset $G/H$ model. In this case one can define also a holomorphic and gauge invariant wave function
\begin{equation}
\chi(A,B) = \int {\cal D}g e^{-L(g,A,B)},
\end{equation}
where
$$
L(g,A,B) = L(g) + {k\over 2 \pi} \int_{\Sigma} d^2 z  {\rm Tr} A_{\bar{z}}
g^{-1}\partial_{z} g 
- {k\over 2 \pi} \int_{\Sigma} d^2 z  {\rm Tr} B_{z}
\partial_{\bar{z}} g  \cdot g^{-1}
$$
\begin{equation}
+ {k \over 2 \pi} \int_{\Sigma}  d^2 z  {\rm Tr}  B_{z} g A_{\bar{z}} g^{-1}
- {k \over 4 \pi} \int_{\Sigma}  d^2 z  {\rm Tr} (A_{z} A_{\bar{z}} +  B_{z}B_{\bar{z}}).
\label{ghmodel}
\end{equation}

These wave function satisfies two copies of the system given by Eqs. (\ref{holo}) and (\ref{gauge}) for both connections $A$ and $B$ with {\it opposite} complex structures. 
These equations implies the existence of a connection on a trivial holomorphic line bundle ${\cal L}^{\otimes (k)}$
over the cartesian product ${\cal C}_J={\cal A}_J \times \overline{\cal B}_J$ of ${\cal A}_J$ the space
of $A$-connections and  ${\cal B}_J$ the space of $B$-connections and the fact that $\psi(A_{\bar{z}},B_z)$ can be regarded as a holomorphic section of this prequantum line bundle. Here $\overline{\cal B}_J$ is ${\cal B}_J$ with the opposite complex structure. The product space ${\cal C}_J={\cal A}_J \times \overline{\cal B}_J$ has the structure of a symplectic
manifold with symplectic structure given by: $\omega = k \omega_0$, where $\omega_0= {1 \over 2 \pi} \int_{\Sigma}
{\rm Tr} (\delta A \wedge \delta A) - {1 \over 2 \pi} \int_{\Sigma}
{\rm Tr} (\delta B \wedge \delta B).$

The corresponding prequantization over this product ${\cal C}_J={\cal A}_J \times \overline{\cal B}_J$ is given by $({\cal L}^{\otimes k}, {\nabla}, \langle \cdot |
\cdot\rangle_{{\cal L}})$, with ${\cal L}^{\otimes (k)}= {\cal
L}^{\otimes (k)}_{(1)} \otimes {\cal L}^{\otimes (-k)}_{(2)}$, where ${\cal
L}^{\otimes (k)}_{(1)}$ is the line bundle over ${\cal A}_J$ and ${\cal L}^{\otimes (k)}_{(2)}$ is the line bundle over 
${\cal B}_J$. Let ${\cal M}_J= {\cal
A}_J/\widehat{G}_{\mathbb{C}}$ and ${\cal N}_J= {\cal B}_J/\widehat{H}_{\mathbb{C}}$ be the moduli spaces and
${\cal R}_J={\cal C}_J/\widehat{F}'_{\mathbb{C}}$ be the quotient space, where $\widehat{F}'_{\mathbb{C}}=\widehat{G}_{\mathbb{C}} \times \widehat{H}_{\mathbb{C}}/Z$. When the group $Z$ is trivial, ${\cal R}_J$ is the product manifold ${\cal R}_J={\cal M}_J \times \overline{\cal N}_J$ which is a K\"ahler manifold and can be regarded as the quotient space of the ${\cal A}_J \times \overline{\cal B}_J$ by the diagonal action of the group $\widehat{G}_{\mathbb{C}} \times \widehat{H}_{\mathbb{C}}$. For the general case of non-trivial $Z$ the pushed-down prequantization with $\widetilde{\cal L} = {\cal
L}^{\widehat{F}'_{\mathbb{C}}},$ being the ${\widehat{F}'_{\mathbb{C}}}$-invariant complex unitary line bundle over ${\cal R}_J.$ For trivial $Z$ we will consider the space of holomorphic sections on $\widetilde{\cal L}_{(1)} \otimes \widetilde{\cal L}^*_{(2)}=  \widetilde{\cal
L}^{\otimes (k)}_{(1)} \otimes \widetilde{\cal L}^{\otimes (-k)}_{(2)}$, which are ${\widehat{F}'_{\mathbb{C}}}$-invariant.  This is given by 
$$
\widetilde{W} = H^0_{L^2}({\cal R}_J, \widetilde{\cal L}^{\otimes k})=
H^0_{L^2}({\cal A}_J \times \overline{\cal B}_J, {\cal L}^{\otimes k})^{\widehat{G}_{\mathbb{C}} \times \widehat{H}_{\mathbb{C}}}
$$
\begin{equation}  
  = H^0_{L^2}({\cal A}_J \times \overline{\cal B}_J, {\cal
L}^{\otimes (k)}_{(1)} \otimes  {\cal L}^{\otimes (-k)}_{(2)})^
{\widehat{G}_{\mathbb{C}} \times \widehat{H}_{\mathbb{C}}}
 = \widetilde{V}_G \otimes \widetilde{V}_{H^*},
\end{equation}
where $ \widetilde{V}_G =  H^0_{L^2}({\cal A}_J, {\cal
L}^{\otimes (k)}_{(1)})^{\widehat{G}_{\mathbb{C}}}$, $\widetilde{V}_{H}=H^0_{L^2}({\cal B}_J, {\cal L}^{\otimes (k)}_{(2)})^{\widehat{H}_{\mathbb{C}}}$ and $\widetilde{V}_{H^*}=H^0_{L^2}(\overline{\cal B}_J, {\cal L}^{\otimes (-k)}_{(2)})^{\widehat{H}_{\mathbb{C}}}$. $\widetilde{V}_{H^*}$ is the dual space to $\widetilde{V}_{H}$. These spaces are precisely the spaces of conformal blocks of the coset models and they are finite dimensional if ${\cal R}$ is compact. For non-trivial $Z$ the space of sections $\widetilde{W}^{G \times H}$ should be modified to take the $Z'$-invariant part $\widetilde{W}^{Z'}=(\widetilde{V}_G \otimes \widetilde{V}_{H^*})^{Z'}$, where $Z'$ arises in the exact sequence:
$0\to i({\widehat{F}_{\mathbb{C}}}) \to {\widehat{F}'_{\mathbb{C}}} \to Z' \to 0$. Then $Z'$ can be defined as the quotient: $Z'= {\widehat{F}'_{\mathbb{C}}}/ i({\widehat{F}_{\mathbb{C}}}),$ where $i: {\widehat{F}_{\mathbb{C}}} \to 
{\widehat{F}'_{\mathbb{C}}}$ is the natural projection map \cite{wzw}.

The inner product $\langle \cdot |\cdot\rangle_{\widetilde{\cal L}}$ on $\widetilde{W}^{Z'}$ is the 
$Z'$-invariant inner product compatible with the
connection $\widetilde{\nabla}$ constructed from $\langle \cdot |
\cdot\rangle_{{\cal L}}$ and it is given by 
$$
\langle \widetilde{\chi} | \widetilde{\psi} \rangle_{\widetilde{{\cal L}}_{(1)} \otimes  \widetilde{{\cal L}}_{(2)}}= \langle \chi 
| \psi
\rangle^{Z'}_{{{\cal L}}_{(1)} \otimes {{\cal L}}_{(2)}} = \langle {\chi} | {\psi} \rangle_
{{{\cal L}}_{(1)} \otimes {{\cal L}}_{(2)}}
$$
\begin{equation}
={1 \over {\rm vol}(\widehat{G}_{\mathbb{C}})
\times {\rm
vol}(\widehat{H}_{\mathbb{C}})} \int_{{\cal A}_J \times \overline{\cal B}_J} {\cal
D}A {\cal D}B \overline{\chi}(A,B) {\psi}(A,B),
\end{equation}
where ${\cal D}A$ and ${\cal D}B$ can be written in terms of the symplectic form $\omega$ it yields:
\begin{equation}
\langle {\chi} | {\psi} \rangle_{{{\cal L}}_{(1)} \otimes {{\cal L}}_{(2)}}
={1 \over {\rm vol}(\widehat{G}_{\mathbb{C}}) \times {\rm
vol}(\widehat{H}_{\mathbb{C}})} \int_{{\cal A}_J \times \overline{\cal B}_J} \langle
\widetilde{\chi} | \widetilde{\psi} \rangle
{\widetilde{\omega}^n \over n!}. 
\end{equation}
By using the previous definitions, the inner product of the tensor product space can be carried over to the form
$$
\langle \widetilde{\chi}(A) \otimes \widetilde{\chi}(B)|\widetilde{\psi}(A) \otimes \widetilde{\psi}(B) \rangle_{\widetilde{{\cal L}}_{(1)} \otimes  \widetilde{{\cal L}}_{(2)}}
$$
$$
=
\bigg({1 \over {\rm vol}(\widehat{G}_{\mathbb{C}})} \int_{{\cal A}_J} {\cal
D}A\overline{\chi}(A) \psi(A)\bigg) \cdot \bigg(
{1 \over {\rm vol}(\widehat{H}_{\mathbb{C}})} \int_{{\cal B}_J} {\cal D}B 
\overline{\chi}(B) {\psi}(B) \bigg)
$$
\begin{equation}
= 
\langle \widetilde{\chi}(A)|\widetilde{\psi}(A)\rangle_{\widetilde{{\cal L}}_{(1)}} \cdot 
\langle \widetilde{\chi}(B)|\widetilde{\psi}(B)\rangle_{\widetilde{{\cal L}}_{(2)}},
\label{factorone}
\end{equation}
for all $\widetilde{\chi}(A,B),\widetilde{\psi}(A,B) \in \big(\widetilde{V}_G \otimes \widetilde{V}_{H^*}\big)^{Z'}$.
Then the inner product factorizes into two pieces corresponding to the group factors $\widehat{G}_{\mathbb{C}}$ and 
$\widehat{H}_{\mathbb{C}}$.
In fact, this is the unique $Z'$-invariant inner product in $\widetilde{W}$ \cite{murphy}.

The norm of an element $\widetilde{\psi}(A,B)$ of $\big(\widetilde{V}_G \otimes \widetilde{V}_{H^*}\big)^{Z'}$
has shown to factorize holomorphically being equal to the partition function of the $G/H$ model \cite{wzw}
$$
Z_{G/H}(\Sigma)= \langle
\widetilde{\psi} |
\widetilde{\psi}
\rangle_{\widetilde{\cal L} \otimes \widetilde{\cal L}^*} \equiv [||\widetilde{\psi}(A) \otimes \widetilde{\psi}(B)
||^2]_{\widetilde{\cal L} \otimes \widetilde{\cal L}^*}
$$
\begin{equation}
={1 \over {\rm vol}(\widehat{G}_{\mathbb{C}})
\times {\rm
vol}(\widehat{H}_{\mathbb{C}})} \int {\cal D}A {\cal D}B {\cal D}g {\cal D}h e^{-L(g,A,B)
-L'(h,A,B)}
\end{equation}
where 
$$
L'(h,A,B) = L(h) + {k\over 2 \pi} \int_{\Sigma} d^2 z  {\rm Tr} B_{\bar{z}}
h^{-1}
\partial_{{z}} h 
- {k\over 2 \pi} \int_{\Sigma} d^2 z  {\rm Tr} A_{z}
\partial_{\bar{z}} h \cdot h^{-1}
$$
\begin{equation}
+ {k \over 2 \pi} \int_{\Sigma}  d^2 z  {\rm Tr}   A_{{z}}h B_{\bar{z}} h^{-1}
- {k \over 4 \pi} \int_{\Sigma}  d^2 z  {\rm Tr} ( A_z  A_{\bar{z}} +  B_{z}B_{\bar{z}}).
\end{equation}
From this it is easy to show that the partition function of the $G/H$ model reduces to the product of factors 
\begin{equation}
Z_{G/H}(\Sigma)=  [||\widetilde{\psi}(\widetilde{A})||^2]_{\widetilde{\cal L}} \cdot [||\widetilde{\psi}(\widetilde{B})
||^2]_{\widetilde{\cal L}^*},
\label{norm}
\end{equation}
corresponding to the group factors of $\widehat{G}_{\mathbb{C}} \times \widehat{H}_{\mathbb{C}}$. Finally one have to take the $Z'$-invariant part.

Removing the zero section we take $\widetilde{\cal Q} \in
\widetilde{\cal L}_{0(1)}^{\otimes (k)} \otimes \widetilde{\cal L}_{0(2)}^{\otimes (-k)},$
$\pi[\widetilde{\cal Q}] = (\widetilde{A},\widetilde{B}) \in {\cal M}_J \times \overline{\cal N}_J$
with local complex
coordinates $\{\widetilde{A}_z,\widetilde{A}_{\bar{z}}; \widetilde{B}_z,\widetilde{B}_{\bar{z}}\}$ and $\pi[\widetilde{\cal Q}'] =
(\widetilde{A}',\widetilde{B}') \in {\cal M}_J \times \overline{\cal N}_J$ with
local complex coordinates $\{\widetilde{A}'_w,\widetilde{A}'_{\bar{w}},\widetilde{B}'_w,\widetilde{B}'_{\bar{w}}\}$. Now
consider the holomorphic section $\widetilde{\psi}(\widetilde{A}_{\overline{z}}, \widetilde{B}_z) = \widetilde{\psi}[\pi(\widetilde{\cal
Q})] =
\widetilde{L_{\cal Q}} [\widetilde{\psi}]
\widetilde{\cal Q}$ with  $\widetilde{L_{\cal Q}} [\widetilde{\psi}]$ being a linear functional of
$\widetilde{\psi}$. 

In the present situation there exists also a section $\widetilde{e}_{\cal Q} =\widetilde{e}^{(1)}_{\cal Q} \otimes \widetilde{e}^{(2)}_{\cal Q} \in  \big(\widetilde{V}_G \otimes \widetilde{V}_{H^*}\big)^{Z'}.$ According to Eq. (\ref{norm}) it has a norm 
\begin{equation}
[||\widetilde{e}_{\cal Q}(A) \otimes \widetilde{e}_{\cal Q}(B)
||^2]_{\widetilde{\cal L} \otimes \widetilde{\cal L}^*}
=[||\widetilde{e}_{\cal Q}(A)||^2]_{\widetilde{\cal L}} \cdot [||\widetilde{e}_{\cal Q}(B)
||^2]_{\widetilde{\cal L}^*}.
\label{factortwo}
\end{equation}

The bounded operator  $\widehat{\cal O}^{Z'}= \widehat{\cal O}^{\widehat{G}_{\mathbb{C}}} \otimes 
\widehat{\cal O}^{\widehat{H}_{\mathbb{C}}} : \big( \widetilde{V}_G \otimes \widetilde{V}_{H^*}\big)^{Z'} \to \big(\widetilde{V}_G \otimes \widetilde{V}_{H^*}\big)^{Z'}$ can be recovered from its symbol in the form $\widehat{\cal O}^{Z'}
\widetilde{\psi} (\widetilde{A}_{\overline{z}},\widetilde{B}_z) = \langle e_{{\cal Q}}| \widehat{\cal O}| \psi
\rangle_{\cal L}^{Z'} {\cal Q}$ this yields
\begin{equation}
\widehat{\cal O} \psi (A_{\overline{z}},B_z) = \widehat{\cal O}^{\widehat{G}_{\mathbb{C}}} \otimes \widehat{\cal O}^{\widehat{H}_{\mathbb{C}}} \bigg(\psi(A_{\overline{z}})
\otimes \psi(B_z)\bigg) = \widehat{\cal O}^{\widehat{G}_{\mathbb{C}}} \psi(A_{\overline{z}}) \otimes \widehat{\cal O}^{\widehat{H}_{\mathbb{C}}} \psi(B_z). 
\end{equation}

The covariant symbol is defined as
\begin{equation}
{\cal O}_B^{Z'}(\widetilde{A}_{\overline{z}}, \widetilde{B}_z) = { \langle \widetilde{e_{\cal Q}}|
\widehat{\cal O}^{Z'}| \widetilde{e_{\cal Q}} \rangle_{\widetilde{\cal L}_{(1)} \otimes \widetilde{\cal L}_{(2)} } \over [||
\widetilde{e_{\cal Q}} ||^2]_{\widetilde{\cal L}_{(1)} \otimes \widetilde{\cal L}_{(2)}}}.
\end{equation}
Using the properties (\ref{factorone}) and (\ref{factortwo}) one can show that
\begin{equation}
{\cal O}_B^{Z'}(\widetilde{A}_{\overline{z}}, \widetilde{B}_z) ={ \langle e_{\cal Q}| \widehat{\cal O}|
e_{\cal Q}
\rangle_{{\cal L}_{(1)} \otimes {\cal L}_{(2)}}^{Z'} 
\over [|| e_{\cal Q} ||^2]_{{\cal L}_{(1)} \otimes {\cal L}_{(2)}}^{Z'}} = 
{ \langle e_{\cal Q}| \widehat{\cal O}|
e_{\cal Q}\rangle_{{\cal L}_{(1)}}^{\widehat{G}_{\mathbb{C}}} 
\over [|| e_{\cal Q} ||^2]_{{\cal L}_{(1)}}^{\widehat{G}_{\mathbb{C}}}} 
\cdot { \langle e_{\cal Q}| \widehat{\cal O}|
e_{\cal Q} \rangle_{{\cal L}_{(2)}}^{\widehat{G}_{\mathbb{C}}}  \over 
[|| e_{\cal Q} ||^2]_{{\cal L}_{(2)}}^{\widehat{H_{\mathbb{C}}}}}.
\label{covasymbol}
\end{equation}
In other worlds it factorizes holomorphically
\begin{equation}
{\cal O}_B^{Z'}(\widetilde{A}_{\overline{z}}, \widetilde{B}_z) = {\cal O}_B^{\widehat{G}_{\mathbb{C}}}(\widetilde{A}_{\overline{z}}) \cdot \overline{\cal O}_B^{\widehat{H}_{\mathbb{C}}}(\widetilde{B}_z),
\label{facsymbol}
\end{equation}
into the product of the two holomorphic symbols ${\cal O}_B^{\widehat{G}_{\mathbb{C}}}(\widetilde{A}_{\overline{z}})$ and $\overline{\cal O}_B^{\widehat{H}_{\mathbb{C}}}(\widetilde{B}_z)$. These symbols correspond to the linear operators: $\widehat{\cal O}^{\widehat{G}_{\mathbb{C}}}: \widetilde{V}_G \to \widetilde{V}_G$ and 
$\overline{\widehat{\cal O}}^{\widehat{H}_{\mathbb{C}}}: \widetilde{V}_{H^*} \to \widetilde{V}_{H^*}.$ 

This implies that the space of covariant symbols $\widetilde{{\cal S}_B}$ is
actually the
tensor product $\big(\widetilde{\cal S}_B^{\widehat{G}_{\mathbb{C}}} \otimes \widetilde{\cal S}_B^{\widehat{H}_{\mathbb{C}}}\big)^{Z'}.$ This tensor product can be analytically 
continued to the open dense subset of
${\cal M}_J \times {\cal M}_J \times \overline{\cal N}_J \times \overline{\cal N}_J$ in such a
way that it can be written
\begin{equation}
{\cal O}_B^{Z'}(\widetilde{A}_{\overline{z}},\widetilde{B}_z;\widetilde{A}'_w,\widetilde{B}'_{\overline{w}}) = {\langle e_{{\cal Q}}| \widehat{\cal
O}| e_{{\cal
Q}'}\rangle_{{\cal L}_{(1)}\otimes {\cal L}_{(2)}}^{Z'} \over
\langle e_{{\cal Q}} | e_{{\cal Q}'}\rangle_{{\cal L}_{(1)} \otimes {\cal L}_{(2)}}^{Z'}}.
\end{equation}
Following a similar procedure as in getting (\ref{covasymbol}) we obtain that the extended symbol also factorizes holomorphically
$$
{\cal O}_B^{Z'}(\widetilde{A}_{\overline{z}},\widetilde{B}_z;\widetilde{A}'_w,\widetilde{B}'_{\overline{w}})  = 
{ \langle e_{\cal Q}| \widehat{\cal O}|
e_{{\cal Q}'}
\rangle_{{\cal L}_{(1)}}^{\widehat{G}_{\mathbb{C}}} 
\over \langle e_{{\cal Q}} | e_{{\cal Q}'}\rangle_{{\cal L}_{(1)}}^{\widehat{G}_{\mathbb{C}}}} \cdot 
{ \langle e_{\cal Q}| \widehat{\cal O}|
e_{{\cal Q}'}
\rangle_{{\cal L}_{(2)}}^{\widehat{H_{\mathbb{C}}}} 
\over 
\langle e_{\cal Q} | e_{{\cal Q}'}\rangle_{{\cal L}_{(2)}^{\widehat{H_{\mathbb{C}}}}}} 
$$
\begin{equation}
= {\cal O}_B^{\widehat{G}_{\mathbb{C}}}(\widetilde{A}_{\overline{z}},\widetilde{A}'_w) \cdot \overline{\cal O}_B^{\widehat{H}_{\mathbb{C}}}(\widetilde{B}_z,\widetilde{B}'_{\overline{w}}),
\label{prodsymb}
\end{equation}
where ${\cal O}_B^{\widehat{G}_{\mathbb{C}}}(\widetilde{A}_{\overline{z}},\widetilde{A}'_w)$ and $\overline{\cal O}_B^{\widehat{H}_{\mathbb{C}}}(\widetilde{B}_z,\widetilde{B}'_{\overline{w}})$ are the corresponding extended symbols. Here we have used the fact that the Bergman kernel ${\cal B}_{\cal Q}(\widetilde{A}_{\overline{z}},\widetilde{B}_z;\widetilde{A}'_w,\widetilde{B}'_{\overline{w}})= \langle \widetilde{e}_{\cal Q} | \widetilde{e}_{{\cal Q}'} \rangle_{\widetilde{\cal L}_{(1)} \otimes \widetilde{\cal L}_{(2)}}$ can be also factorized as $\langle \widetilde{e}_{\cal Q}^{(1)} | \widetilde{e}_{{\cal Q}'}^{(1)} \rangle_{\widetilde{\cal L}_{(1)}} \cdot \langle \widetilde{e}_{\cal Q}^{(2)} | \widetilde{e}_{{\cal Q}'}^{(2)} \rangle_{\widetilde{\cal L}_{(2)}}$ and consequently
\begin{equation}
{\cal B}_{\cal Q}(\widetilde{A}_{\overline{z}},\widetilde{B}_z;\widetilde{A}'_w,\widetilde{B}'_{\overline{w}}) = {\cal B}_{\cal Q}(\widetilde{A}_{\overline{z}},\widetilde{A}'_w) \cdot \overline{\cal B}_{\cal Q}(\widetilde{B}_z,\widetilde{B}'_{\overline{w}}).  
\label{factorB}
\end{equation}

Now if ${\cal O}_{B(0)}(\widetilde{A}_{\overline{z}}, \widetilde{B}_z;\widetilde{A}'_{{w}},\widetilde{B}'_{\overline{w}})$ and ${\cal
O}'_{B(0)}(\widetilde{A}'_{\overline{w}},\widetilde{B}'_w;\widetilde{A}_{{z}},\widetilde{B}_{\overline{z}})$ are two covariant
symbols of $\widehat{\cal O}_0$ and $\widehat{{\cal O}'}_0$, respectively, then the
covariant symbol of $\widehat{\cal O}_0\widehat{{\cal O}'}_0$ is given by the 
Berezin-Wick star product
$$
({\cal O}^{(j)}_{B(0)} \widetilde{*_B} {\cal O}'^{(j)}_{B(0)})(\widetilde{A}_{\overline{z}},\widetilde{B}_z;\widetilde{A}_{{z}},\widetilde{B}_{\overline{z}})
$$
$$
=\int_{{\cal W}^{(j)}_J \times \overline{\cal W}^{(j)}_J} {\cal O}^{(j)}_{B(0)}(\widetilde{A}_{\overline{z}},\widetilde{B}_z;\widetilde{A}'_{{w}},\widetilde{B}'_{\overline{w}})
{\cal
O}'^{(j)}_{B(0)}(\widetilde{A}'_{\overline{w}},\widetilde{B}'_w;\widetilde{A}_{{z}}, \widetilde{B}_{\overline{z}}) 
$$
$$
\times  {{\cal B}^{(j)}(\widetilde{A}_{\overline{z}},\widetilde{B}_z;\widetilde{A}'_{{w}},\widetilde{B}'_{\overline{w}}){\cal B}^{(j)}(\widetilde{A}'_{\overline{w}},\widetilde{B}'_w;\widetilde{A}_{{z}},\widetilde{B}_{\overline{z}}) \over {\cal
B}^{(j)}(\widetilde{A}_{\overline{z}},\widetilde{B}_z;\widetilde{A}_{{z}},\widetilde{B}_{\overline{z}})}
\exp \bigg\{- \Phi^{(j)}(\widetilde{A}'_{\overline{w}},\widetilde{B}'_w;\widetilde{A}'_{{w}},\widetilde{B}'_{\overline{w}}) \bigg\} {\widetilde{\omega} \over
n!}(\widetilde{A}'_{\overline{w}},\widetilde{B}'_w;\widetilde{A}'_{{w}},\widetilde{B}'_{\overline{w}})
$$
$$
= \int_{{\cal W}^{(j)}_J \times \overline{\cal W}^{(j)}_J } {\cal O}^{(j)}_{B(0)}(\widetilde{A}_{\overline{z}},\widetilde{B}_z;\widetilde{A}'_{{w}},\widetilde{B}'_{\overline{w}}) 
{\cal O}'^{(j)}_{B(0)}(\widetilde{A}'_{\overline{w}},\widetilde{B}'_w;\widetilde{A}_{{z}}, \widetilde{B}_{\overline{z}})
$$
\begin{equation}
\times 
\exp \bigg\{
{\cal K}^{(j)}(\widetilde{A}_{\overline{z}},\widetilde{B}_z;\widetilde{A}_{{z}},\widetilde{B}_{\overline{z}} |\widetilde{A}'_{\overline{w}},\widetilde{B}'_w;\widetilde{A}'_{{w}},\widetilde{B}'_{\overline{w}}) \bigg \} {\widetilde{\omega} \over n!}
(\widetilde{A}'_{\overline{w}},\widetilde{B}'_w;\widetilde{A}'_{{w}},\widetilde{B}'_{\overline{w}}) 
\end{equation}
where ${\cal K}^{(j)}(\widetilde{A}_{\overline{z}},\widetilde{B}_z;\widetilde{A}_{{z}},\widetilde{B}_{\overline{z}}|\widetilde{A}'_{\overline{w}},\widetilde{B}'_w;\widetilde{A}'_{{w}},\widetilde{B}'_{\overline{w}}):=
\Phi^{(j)}(\widetilde{A}_{\overline{z}},\widetilde{B}_z;\widetilde{A}'_{{w}},\widetilde{B}'_{\overline{w}})+
\Phi^{(j)}(\widetilde{A}'_{\overline{w}},\widetilde{B}'_w;\widetilde{A}_{{z}},\widetilde{B}_{\overline{z}})-
\Phi^{(j)}(\widetilde{A}_{\overline{z}},\widetilde{B}_z;\widetilde{A}_{{z}},\widetilde{B}_{\overline{z}})- \Phi^{(j)}(\widetilde{A}'_{\overline{w}},\widetilde{B}'_w;\widetilde{A}'_{{w}},\widetilde{B}'_{\overline{w}})$ is called the Calabi
diastatic
function on ${\cal W}^{(j)}_J \times \overline{\cal W}^{(j)}_J$. This construction is valid for all local prequantization
$(\widetilde{\cal L}_{(j)}^{\otimes k}, \widetilde{\nabla}^{(j)}, \langle \cdot |
\cdot\rangle_{\widetilde{\cal
L}_{(j)}})$. Finally, this structure leads to the pair $(\widetilde{{\cal S}_B}, \widetilde{*_B})$ which
constitutes the Berezin
quantization of $({\cal R}_J,\widetilde{\omega})$. It an easy matter to see, using all previous
results, that Berezin product also factorizes holomorphically 
\begin{equation}
({\cal O}^{(j)}_{B(0)} \widetilde{*_B} {\cal O}'^{(j)}_{B(0)})(\widetilde{A}_{\overline{z}},\widetilde{B}_z;
\widetilde{A}_z,\widetilde{B}_{\overline{z}}) = \bigg({\cal O}^{(j)}_{B(0)} *_B {\cal O}^{(j)}_{B(0)}\bigg)(\widetilde{A}_z,\widetilde{A}_{\overline{z}})
\cdot \bigg(\overline{{\cal O}'^{(j)}_{B(0)} *_B {\cal O}'^{(j)}_{B(0)}}\bigg)( \widetilde{B}_z,\widetilde{B}_{\bar{z}}).
\end{equation}

\subsection{The $G/G$ Model}

In the present subsection we specialize the discussion in the previous subsection to the case $\widehat{H}_{\mathbb{C}}=\widehat{G}_{\mathbb{C}}$. In this case very
interesting features about the gauged coset model $G/G$ arises. First af all this case corresponds to a topological quantum field theory discoverd by Witten in \cite{wzw} and applied in \cite{grassman}. The `anomalous' Lagrangian is basically given by Eq. (\ref{ghmodel}) with the addition that now $B$ is like $A$ a $G$-valued connection. In this case Lagrangian (\ref{ghmodel})
has the additional symmetry: 
\begin{equation}
z \leftrightarrow \overline{z},  \ \ \ \ \  A \leftrightarrow B,  \ \ \ \ \   g \leftrightarrow g^{-1}.
\end{equation}
Thus, complex conjugate $\overline{\chi}(A,B)$ of $\chi(A,B)$ can be computed leaving the complex structure fixed but changing $A \leftrightarrow B$ and $g \leftrightarrow g^{-1}$. Remember that,
$\widetilde{\chi}(\widetilde{A},\widetilde{B})= \widetilde{\chi}(\widetilde{A}) \otimes \widetilde{\chi}(\widetilde{B}) \in \big(\widetilde{V}_G \otimes \widetilde {V}_{H^*}\big)^{Z'}.$ In our present case 
$\widehat{H}_{\mathbb{C}}=\widehat{G}_{\mathbb{C}},$  we have $\widetilde{\chi}(A,B)= \widetilde{\chi}(A) \otimes \widetilde{\chi}(B) \in  \big(\widetilde{V}_G \otimes \widetilde {V}_{G^*}\big)^{Z'} = \bigg({\rm Hom}\big(\widetilde{V}_G,\widetilde{V}_G \big)\bigg)^{Z'}$. This fact and the symmetry $A \leftrightarrow B$ implies that
$\widetilde{\chi}(A,B)$ is Hermitian, {\it i.e.} $\widetilde{\chi}(A,B)= \overline{\widetilde{\chi}(B,A)}$. Also it satisfies the property:
$\widetilde{\chi}^2 = \widetilde{\chi},$ which corresponds to an orthogonal projector of the bundle ${\cal V}$ onto a finite rank sub-bundle ${\cal V}'$. This implies that for any finite, holomorphic and orthonormal basis $\{{\bf e}_i(A;\rho)\}$ of ${\cal V}'$, any section $ \chi$ can be written diagonally as 
\begin{equation}
\chi(A,B;\rho) =\sum_{i,j=1}^{{\rm dim} {\cal V}'}  \delta_{ij} {\bf e}_i(A;\rho) \otimes \overline{{\bf e}}_j(B;\rho).
\end {equation}
By the usual rules of tensor products of Hilbert spaces \cite{murphy} we have the topological invariant
\begin{equation}
Z_{G/G}(\Sigma) = |\chi|^2 = {\rm dim} ({\cal V}'),
\end{equation}
where ${\cal V}'$ is the space of conformal blocks of the WZW model.  

The generalized coherent states can be also expressed in this basis depending on the complex structure 
\begin{equation}
\widetilde{e_{\cal Q}}(\widetilde{A}_{\overline{z}},\widetilde{B}_z; \rho) = \sum_{i=1}^{{\rm dim} {\cal V}'} {\bf e}_i(\widetilde{A}_{\overline{z}};\rho) \otimes \overline{\bf e}_i(\widetilde{B}_z;\rho).
\label{basedos}
\end{equation}
It is easy to see that the norm of $e_{\cal Q}(A,B; \rho)$ is given by
\begin{equation}
[||\widetilde{e}_{\cal Q}(A;\rho) \otimes \widetilde{e}_{\cal Q}(B;\rho)
||^2]_{{\rm Hom}(\widetilde{\cal L}, \widetilde{\cal L})}
=[||\widetilde{e}_{\cal Q}(A;\rho)||^2]_{\widetilde{\cal L}} \cdot [||\widetilde{e}_Q(B;\rho)
||^2]_{\widetilde{\cal L}^*} = {\rm dim} {\cal V}'.
\end{equation}

The covariant symbol is defined as before but taking $\widehat{H}_{\mathbb{C}} = \widehat{G}_{\mathbb{C}}$ in Eq. (\ref{covasymbol}). After some computations we finally get
\begin{equation}
{\cal O}_B^{Z'}(\widetilde{A}_{\overline{z}}, \widetilde{B}_z;\rho) ={1 \over ({\rm dim} {\cal V}')^2} \sum_{k, \ell =1}^{{\rm dim}{\cal V}'} {\cal O}_{B \  k \ell}^{\widehat{G}_{\mathbb{C}}}(\widetilde{A}_{\overline{z}};\rho) \overline{\cal O}_{B  \ell k}^{\widehat{G}_{\mathbb{C}}}(\widetilde{B}_z;\rho),
\end{equation}
which coincides with the holomorphic factorization (\ref{facsymbol}). The space of covariant symbols $\widetilde{{\cal S}_B}$ is
actually the
tensor product $\big(\widetilde{\cal S}_B^{\widehat{G}_{\mathbb{C}}} \otimes (\widetilde{\cal S}_B^{\widehat{G}_{\mathbb{C}}})^*\big)^{Z'}.$ This is isomorphic to the space $\bigg({\rm Hom}\big(\widetilde{\cal S}_B^{\widehat{G}_{\mathbb{C}}}, \widetilde{\cal S}_B^{\widehat{G}_{\mathbb{C}}}\big)\bigg)^{Z'}$, the space of linear matrices ${\rm dim} {\cal V}' \times {\rm dim} {\cal V}'$. This space can be analytically 
continued in order to define the extended symbols. They also satisfy the holomorphic factorization condition 
\begin{equation}
{\cal O}_B^{Z'}(\widetilde{A}_{\overline{z}},\widetilde{B}_z;\widetilde{A}'_w,\widetilde{B}'_{\overline{w}})
= 
{\cal O}_B^{\widehat{G}_{\mathbb{C}}}(\widetilde{A}_{\overline{z}},\widetilde{A}'_w) \cdot \overline{\cal O}_B^{\widehat{G}_{\mathbb{C}}}(\widetilde{B}_z,\widetilde{B}'_{\overline{w}}),
\label{factextsymb}
\end{equation}
where ${\cal O}_B^{\widehat{G}_{\mathbb{C}}}(\widetilde{A}_{\overline{z}},\widetilde{A}'_w)$ and $\overline{\cal O}_B^{\widehat{G}_{\mathbb{C}}}(\widetilde{B}_z,\widetilde{B}'_{\overline{w}})$ are the corresponding extended symbols.

The Bergmann kernel for the $G/G$ model can be written in the basis (\ref{basedos}) as follows
\begin{equation}
{\cal B}_{\cal Q}(\widetilde{A}_{\overline{z}},\widetilde{B}_z;\widetilde{A}'_w,\widetilde{B}'_{\overline{w}}) = \bigg(\sum_{k=1}^{{\rm dim}{\cal V}'} {\bf e}_k(\widetilde{A}_{\overline{z}};\rho) {\bf e}_k(\widetilde{A}'_w;\rho) \bigg) \cdot  \bigg( 
\sum_{\ell=1}^{{\rm dim}{\cal V}'} \overline{\bf e}_{\ell}(\widetilde{B}_z;\rho) \overline{\bf e}_{\ell}(\widetilde{B}'_{\overline{w}};\rho) 
\bigg).
\end{equation}
According to this result and the factorization of the extended symbols (\ref{factextsymb}) we finally get the holomorphic factorization of the Berezin product
\begin{equation}
({\cal O}^{(j)}_{B(0)} \widetilde{*_B} {\cal O}'^{(j)}_{B(0)})(\widetilde{A}_{\overline{z}},\widetilde{B}_{{z}};
\widetilde{A}_z,\widetilde{B}_{\overline{z}}) = \bigg({\cal O}^{(j)}_{B(0)} *_B {\cal O}^{(j)}_{B(0)}\bigg)(\widetilde{A}_{\overline{z}},\widetilde{A}_{{z}})
\cdot \bigg(\overline{{\cal O}'^{(j)}_{B(0)} *_B {\cal O}'^{(j)}_{B(0)}}\bigg)( \widetilde{B}_z,\widetilde{B}_{\bar{z}}).
\end{equation}

\section{Final Remarks}
In this paper we have applied the Berezin quantization global procedure to the gauged WZW and coset models. 
Our description has been formal. For the gauged $G/H$ model we have found that holomorphic factorization of the partition function of the corresponding model can be carried over to the Berezin quantization procedure. Covariant symbols, extended covariant symbols and the Berezin-Wick star product factorizes into two pieces corresponding to the group factors
$\widehat{G}_{\mathbb{C}} \times \widehat{H}_{\mathbb{C}}$. For the topological $G/G$ model, its corresponding Berezin quantization leads to the quantization of the space of linear matrices ${\rm dim} {\cal V}' \times {\rm dim} {\cal V}'$ on the space of conformal blocks of the associated CFT of the $G/G$ model.

It would be interesting to extend the cases considered in this paper to ${\cal N}=1$ and ${\cal
N}=1$ supersymmetric cosets models and to topological Kazama-Suzuki models and its coupling to
topological gravity \cite{Nmatrix,mblau}. It would be equally interesting to see if the Berezin quantization procedure is also applicable to string theory in the version of \cite{fs}. Finally, the application of some recents results 
\cite{isidrodos} to the quantization of gauged WZW and coset models deserves further study. 

\break
\centerline{\bf Acknowledgments}
I. C.-I and H. G.-C. are deeply indebted to Prof. Egidio Barrera, who was a great teacher and friend for us. We would have 
been pleased to have his comments on the matters considered in this paper. Discussions with G. Dito, M. Przanowski and F. Turrubiates are greatly appreciated. This work was supported in part by CONACyT M\'exico Grant
41993F. The work of W. H.-S. is supported by a CONACyT graduate fellowship.


\end{document}